# Gate-tunable Spin-Galvanic Effect in Graphene-Topological insulator van der Waals Heterostructures at Room Temperature

Dmitrii Khokhriakov[1], Anamul Md. Hoque[1], Bogdan Karpiak[1], Saroj P. Dash[1*]

[1]*Department of Microtechnology and Nanoscience, Chalmers University of Technology, SE-41296, Göteborg, Sweden*

## Abstract

Unique electronic spin textures in topological states of matter are promising for emerging spin-orbit driven memory and logic technologies. However, there are several challenges related to the enhancement of their performance, electrical gate-tunability, interference from trivial bulk states, and heterostructure interfaces. We address these challenges by integrating two-dimensional graphene with a three-dimensional topological insulator (TI) in van der Waals heterostructures to take advantage of their remarkable spintronic properties and engineer proximity-induced spin-charge conversion phenomena. In these heterostructures, we experimentally demonstrate a gate-tunable spin-galvanic effect (SGE) at room temperature, allowing for efficient conversion of a non-equilibrium spin polarization into a transverse charge current. Systematic measurements of SGE in various device geometries via a spin switch, spin precession, and magnetization rotation experiments establish the robustness of spin-charge conversion in the Gr-TI heterostructures. Importantly, using a gate voltage, we reveal a strong electric field tunability of both amplitude and sign of the spin-galvanic signal. These findings provide an efficient route for realizing all-electrical and gate-tunable spin-orbit technology using TIs and graphene in heterostructures, which can enhance the performance and reduce power dissipation in spintronic circuits.

**Keywords:** Graphene, Topological Insulator, van der Waals heterostructures, Proximity effect, spin-orbit interaction, Spin Hall effect, Spin-Galvanic effect, Rashba-Edelstein effect.

## Introduction

Spin-based memory, oscillators, and logic devices operating at high speed and low power are promising for future artificial intelligence and information technology[1–3]. In these device architectures, spin-polarized currents are used to exert a spin-orbit torque (SOT) on an adjacent magnetic layer[4], enabling a manipulation, persistent oscillation, and even switching of the magnetization[5,6]. The key to these technologies so far is the use of heavy metals, semiconductors, and heterostructure interfaces of metals and oxides, which allow for spin-charge interconversion due to their strong spin-orbit interaction (SOI) and broken inversion symmetry[7]. Recently, such spin-charge conversion process and its inverse phenomenon were investigated in Weyl semimetals[8,9], Rashba materials[10], transition metal dichalcogenides (TMDs)[11], and graphene/TMD heterostructures[12–14].

To further enhance the spin-charge conversion efficiency and to reduce the power consumption beyond the capacity of conventional materials, topological insulators (TIs) can be introduced[15]. This allows taking advantage of their topologically protected Dirac surface states with a distinct spin-momentum locking (SML) feature, which gives rise to a spontaneous electron spin polarization by application of an electric field. Recently, electric detection of SML states[16–20] and large spin-charge conversion with an efficient SOT[21–23] has been demonstrated up to room temperature using TIs in contact with ferromagnetic materials. In particular, the higher spin-charge conversion efficiency of highly-conducting p-type TIs has been revealed[1,24]. However, there are several challenges related to the electrical tunability of spin-charge conversion in TIs, their unintentional doping issues[25], and possible destruction of topological surface states in contact with ferromagnets at the interface[26].

Graphene-based heterostructures are a particularly exciting device concept since they allow to utilize a strong gate-tunability of graphene (Gr) to study proximity effects arising from its hybridization with other functional materials. Combining graphene with TIs in van der Waals heterostructures is theoretically predicted to introduce a strong SOI with Rashba spin-splitting in the graphene[27–30] while preserving the topological bands of TIs[31]. As a consequence, proximitized graphene becomes a host to the spin-galvanic effect (SGE), also known as the inverse Rashba-Edelstein effect (IREE), which involves the conversion of non-equilibrium spin density into a charge current due to the spin texture in a material[32].

Here, we introduce an atomically-thin graphene layer in van der Waals heterostructures with a p-type TI (($Bi_{0.15}Sb_{0.85})_2Te_3$) to experimentally demonstrate a gate-tunable spin-galvanic effect at room temperature. Measurements of the SGE in different device geometries via a spin switch, Hanle spin precession, and magnetization rotation methods reveal a good agreement with the expected behavior of a proximity-induced Rashba spin texture in the hybrid Gr-TI bands. The observation of gate tunability and sign-switch functionality of the spin-galvanic signal demonstrate an all-electrical operation of a hybrid spintronic device.

# Results

## Spin-galvanic effect detection via spin precession

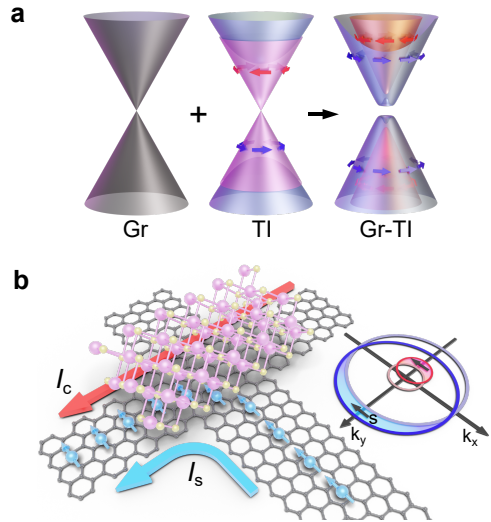

**Figure 1. Spin-galvanic effect in a graphene-topological insulator heterostructure. a,** Schematics of band structures of pristine graphene (Gr), topological insulator (TI), and a Gr-TI heterostructure. Due to proximity-induced spin-orbit interaction (SOI), graphene develops Rashba spin-split bands and acquires a spin texture that is different from the spin texture of the TI surface states. **b**, A schematic representing the spin-galvanic effect, where spin-polarized carriers diffuse in the Gr-TI heterostructure, acquire a transverse momentum and produce a charge current. The inner and outer Fermi circles of spin-split proximitized graphene bands shift in opposite directions in the k-space, providing the carriers with a net momentum along a direction perpendicular to their spin.

Strong proximity-induced SOI in a Gr-TI heterostructure can result in the Rashba spin-splitting of the graphene bands (schematically shown in Fig. 1a), which introduces a spin texture to the graphene. Figure 1b illustrates the principle of SGE, showing the spins polarized along the negative x-direction $s_{-x}$ diffusing as a spin current $I_s$ into the Gr-TI heterostructure region, where the only available states for such spins are those with the momentum $k$ along the y-direction. As the proximitized graphene bands accommodate these spin-polarized carriers, their Fermi circles shift, and asymmetric spin-flip scattering[32] provides carriers with a net momentum along the +y-axis and thus creates a transverse charge current $I_c$ in the heterostructure region. To investigate such spin-charge conversion in a hybrid Gr-TI system, we conceived two types of device geometries, where the required $s_{-x}$ polarization is achieved either by aligning an injector ferromagnet (FM) easy magnetization axis along the x-direction (device 1) or by rotating the FM magnetization with the external magnetic field (devices 2, 3 and 4).

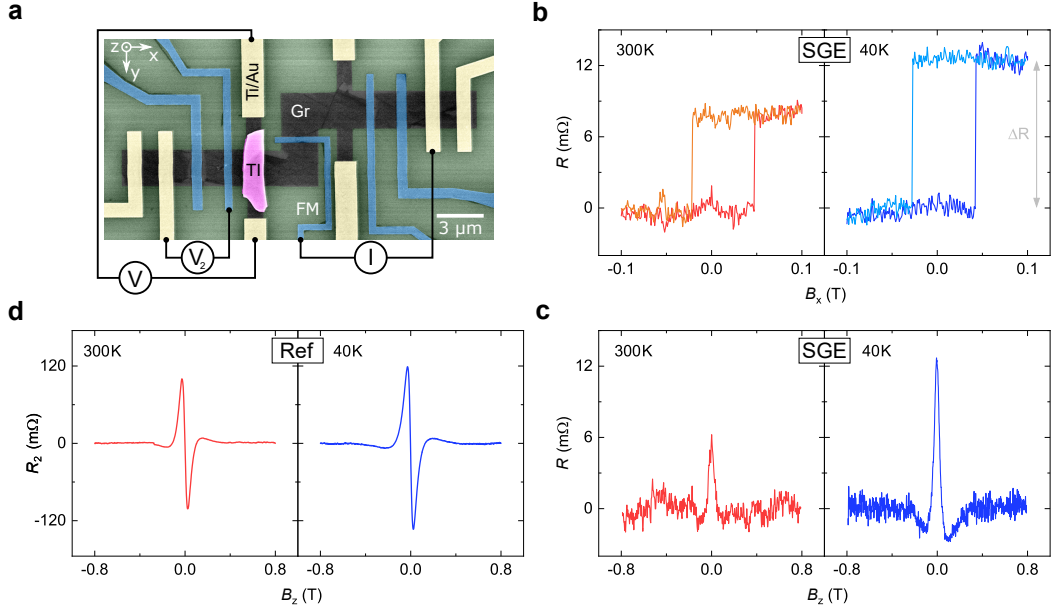

**Figure 2. Measurements of the spin galvanic effect in a graphene-topological insulator heterostructure. a,** A colored scanning electron microscopy picture of a hybrid device consisting of a modified Gr Hall bar structure with a TI flake placed on top of one of the crosses. The device is contacted by ferromagnetic (FM) tunnel electrodes (TiO$_2$/Co, blue) and non-magnetic electrodes (Ti/Au, yellow). The nonlocal measurement configuration includes the spin current injection from a FM into graphene and the SGE detection as a voltage signal $V$ across the Gr-TI Hall cross by non-magnetic contacts. In addition, a reference spin voltage $V_2$ is detected by a second ferromagnet. **b,** The SGE signal $R = V/I$ measured with the in-plane magnetic field $B_x$ at $T$ = 300 K (red) and $T$ = 40 K (blue). **c.** The SGE detected via spin precession with the out-of-plane field $B_z$. **d,** Reference Hanle spin precession measurements $R_2 = V_2/I$ obtained with a ferromagnetic detector. All the measurements were performed with the back gate voltage $V_g$ = -80 V and bias current $I$ = -500 µA.

First, we present results from device 1 with a special geometry (Fig. 2a), consisting of the CVD graphene patterned in a modified Hall bar shape with a flake of TI placed on top of a Hall cross. Single crystal flakes of the TI were exfoliated and transferred onto the graphene inside a glovebox in a controlled environment to achieve clean van der Waals interfaces. The graphene was subsequently patterned and contacted by ferromagnetic (FM, TiO$_2$/Co with contact resistance of 1-3 kΩ) and non-magnetic (Ti/Au with contact resistance of around 1 kΩ) electrodes in a multi-step electron beam lithography process (see Methods). In this device 1, the FM contact used for spin injection is placed perpendicularly to the heterostructure Hall cross. The advantage of such a configuration is that it allows us to perform conclusive SGE measurements via both a spin switch and Hanle spin precession at low magnetic fields, which is not possible with conventional geometries.

In the nonlocal Hall measurement geometry, an electric current $I$ is injected from a FM contact into the graphene channel, creating non-equilibrium spin accumulation polarized along the x-direction. As the diffusive spin current propagates into the Gr-TI region, carriers acquire a transverse momentum and create a charge current along the y-direction due to SGE, which is detected as the voltage *V* across the Gr-TI Hall cross by non-magnetic contacts. Measurements were performed at room temperature (300 K), and at *T* = 40 K. Figure 2b shows the measured SGE signal *R* = *V*/*I*, where sharp changes in the detected nonlocal resistance are observed as the in-plane magnetic field $B_x$ switches the magnetization state of the injector FM electrode. To further establish the SGE, we measured the signal with the application of the out-of-plane field $B_z$, which causes the injected spins to precess and dephase, resulting in a characteristic Hanle shape, as shown in Fig. 2c. The presence of both the spin-switch and Hanle spin precession signals confirms the observation of the SGE in the Gr-TI heterostructure.

A reference spin signal $R_2$ = $V_2$/*I* is detected simultaneously to SGE by the second FM electrode, where an antisymmetric Hanle spin precession signal is observed (Fig. 2d) because of the perpendicularly placed injector and detector ferromagnets[33]. As the device is cooled down from 300 K to 40 K, the magnitude of the reference spin signal increases by a factor of 1.3, while the SGE signal nearly doubles. Assuming that the increase in the reference signal is governed by the increase in spin polarization of both the injector and detector FM contacts, we can conclude that the efficiency of spin to charge conversion by SGE also increases at low temperatures.

To further investigate the nature of the SGE, we performed systematic measurements with varying spin injection bias current *I* applied to the FM injector contact (see Supplementary Figures 1,2). The signals produced by the SGE and the reference spin transport measurements both have an asymmetric dependence on the bias *I*, as the switching occurs in the same direction for both +*I* and -*I* (see Supplementary Note 1). Such behavior is commonly observed in graphene spin transport experiments with tunneling contacts and is generally assigned to magnetic proximity effects and energy-dependent spin-resolved density of states at the injector FM/Gr interface[34]. As the bias trends of the SGE and the reference spin signal show a pronounced correlation, the behavior of both signals can be attributed to the bias-dependent properties of the FM injector contact, while the SGE seems to scale linearly with the value of injected spin polarization. By fitting the Hanle equation to both the SGE and reference spin precession data, we evaluated spin lifetime $\tau_s$ = 150 - 190 ps and spin diffusion length $\lambda_s$ = 2.5 - 3.5 µm of our system (see Supplementary Figures 2e,f). Although the precision of parameter extraction is affected by the non-uniform length of the channel, a good correspondence between the parameters obtained by SGE and reference Hanle measurements allows for possible applications of Gr-TI heterostructures for non-magnetic creation/detection of spin polarization, as well as the characterization of spintronic properties in devices free from ferromagnets.

## Gate – tunability of spin galvanic effect.

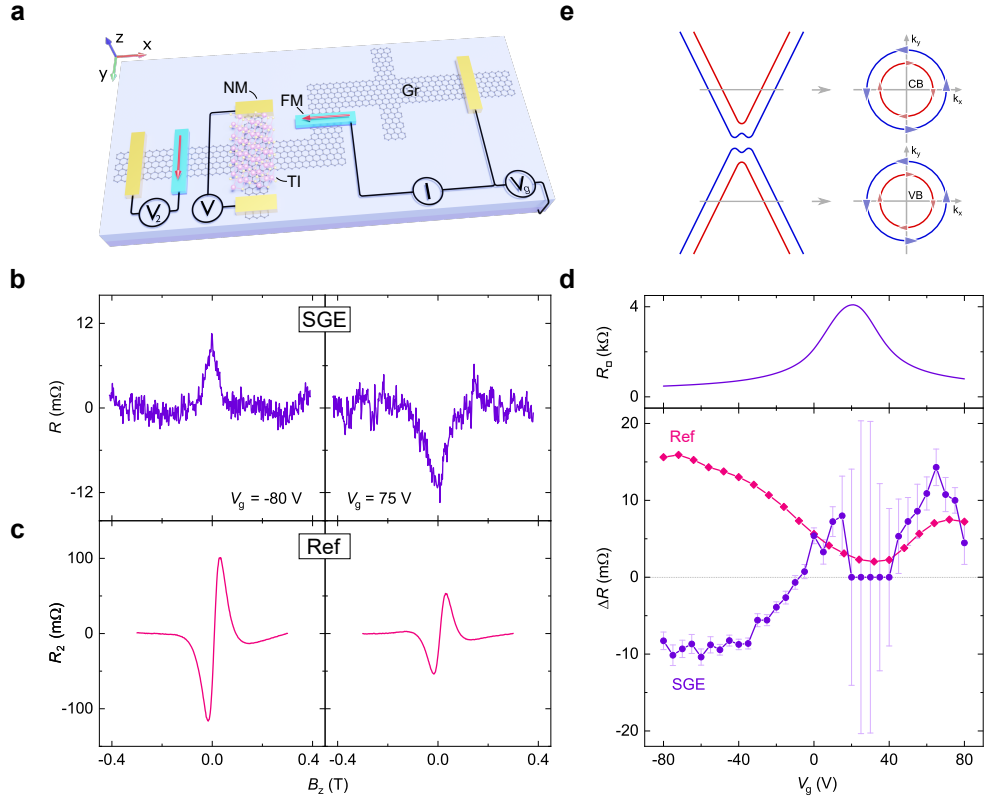

**Figure 3. Gate-controlled sign change of the spin-galvanic signal at room temperature. a,** A schematic of the device and the measurement geometry. **b,** SGE spin precession signals measured with the Fermi level tuned to the conduction ($V_g$ = 75 V) and valence ($V_g$ = -80 V) bands of the proximitized graphene. **c,** Reference nonlocal Hanle measurements obtained with the ferromagnetic detector at $V_g$ = ±80 V. **d,** Lower panel: The amplitudes of the SGE signal (purple) and scaled (x 0.1) reference Hanle signal (pink) as a function of the gate voltage $V_g$. All the measurements were performed at room temperature with the bias current $I$ = -500 µA. Upper panel: Gr-TI channel sheet resistance as a function of the gate voltage, showing a maximum at the charge neutrality point $V_{CNP}$ = 20 V. **e,** A schematic band structure of Gr in proximity to a TI with indicated spin textures of the Rashba spin-split conduction (CB) and valence (VB) bands. Colors represent the helicity of the bands.

One of the most essential functionalities of spintronic devices is the possibility to control the spin information by a gate electric field. An application of the back gate voltage $V_g$ (Fig. 3a) is expected to mainly tune the Fermi level ($E_f$) in the proximitized and pristine graphene regions, while the metallic TI Dirac surface states and bulk bands may be gate-insensitive due to their strong p-type doping. Figure 3b shows the SGE spin precession signals measured at two representative gate voltages, corresponding to the graphene $E_f$ tuned to the conduction ($V_g$ = 75 V) and valence bands ($V_g$ = -80 V). An apparent change of the signal sign is observed, demonstrating gate-tunability and switching functionality of spin-charge conversion in the Gr-TI heterostructures. As the spin precession signals in the reference geometry $V_2$ have the same sign at all gate voltages (Fig. 3c), we can rule out a trivial case of the change in the polarity of injected spins by the gate voltage as the reason for the observed sign change of the SGE. Figure 3d shows the $V_g$ dependence of the magnitudes of SGE, reference Hanle signal, and the channel sheet resistivity $R_\square$. As the graphene is tuned towards its maximal resistivity state, which defines its average charge neutrality point (CNP) $V_{CNP}$ = 20 V, the reference spin signal amplitude shows a minimum, while the SGE signal becomes obscured due to a substantial increase in the noise level of the measurements. Both these features are common to our graphene-based spintronic devices and are believed to arise due to the varying degree of conductivity mismatch between the FM tunnel injector contact and the graphene as the gate voltage tunes the resistivity of the channel. On the other hand, the presence of the sign change point in SGE ($V_0$) between -10 and -5 V and its deviation from $V_{CNP}$ is an interesting feature of the Gr-TI heterostructures that we discuss below.

As predicted by theoretical calculations, the proximity-induced SOI in graphene can give rise to Rashba spin-splitting of the graphene bands, opening a bandgap and inducing a spin texture with the same spin winding direction in low-energy conduction and valence bands[30] (Fig. 3e). Since the band chirality is the same regardless of the carrier type, the electrons and holes with the same spin acquire the momentum in the same direction, which creates the charge current of opposite signs due to the negative (positive) electric charge of electrons (holes). Therefore, the spin-galvanic signal is expected to change sign, but only when the $E_f$ is tuned across the CNP of graphene. However, SGE occurs only in a Gr-TI heterostructure and therefore is expected to reflect only the carrier type change in the proximitized graphene region, which may occur at a gate voltage different from the $V_{CNP}$ obtained by the $R_\square$ measurement since the latter also includes contributions from the pristine graphene areas, as well as the areas of graphene below the electrodes, which may cause additional doping.

Besides SGE, we consider contributions from other spin-charge conversion processes possible in systems with strong SOI. First, we evaluate the contribution from the inverse spin Hall effect (ISHE) in graphene. While the SGE is a 2D effect describing the conversion of spin density into charge current, the ISHE is 3D and describes the conversion of spin current $\mathbf{I}_s$ into charge current $\mathbf{I}_c$ requiring the orthogonality between $\mathbf{I}_s$, $\mathbf{I}_c$ and spin $\mathbf{s}$.[7] Due to this constraint, the ISHE in graphene can only occur for spins polarized out-of-plane with $\mathbf{I}_s$ and $\mathbf{I}_c$ being in plane and perpendicular to each other. In our measurement geometry, spin polarization along the z-axis cannot be created by $B_x$ or spin precession in $B_z$, therefore, all the presented measurements are not sensitive to the proximity-induced ISHE. On the other hand, $s_z$ can be created due to spin precession in $B_y$, which can result in an antisymmetric ISHE signal superimposed with the symmetric Hanle shape of SGE. However, no antisymmetric component could be distinguished in the measured data (Supplementary Figure 3), presumably due to relatively small efficiency of ISHE compared to SGE in our Gr-TI system. We note that this observation is different from the reports on Gr-TMDC systems, where amplitudes of proximity-induced SHE and SGE were found to be comparable[13,35].

Next, we consider possible contributions to the measured signal originating from the TI. Since we use a highly doped metallic TI in van der Waals contact with the graphene (see Supplementary Figure 4 for TI characterization), it can absorb spin current from the channel and contribute with additional spin-charge conversion processes due to the spin textures in its Dirac surface states and Rashba bulk bands. As the TI surface spin texture is opposite for holes and electrons (Fig. 1a), SGE in these states it is not expected to produce a sign change in the measurement[16,19]. Although the TI Rashba spin-split bulk states can yield opposite signs of the SGE in the conduction and valence bands, in our experimental conditions the $E_f$ in the TI is not expected to reach the bulk conduction band due to its high hole doping, and, therefore, we can rule out these states as the origin of observed sign change. Similarly, the ISHE in the TI bulk bands is not expected to contribute to the sign change, since we are not able to change the carrier type in these bands by the gate voltage (see Supplementary Note 2 for further discussion). This leaves the SGE in proximitized graphene as the only mechanism that can be responsible for the observed sign change of the measured spin-charge conversion signal.

**Spin galvanic effect detection by magnetization rotation.**

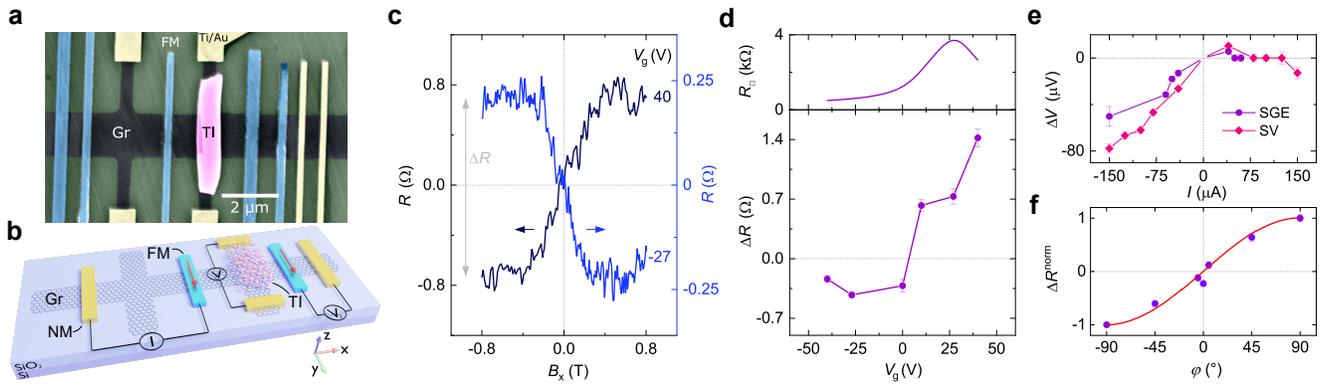

**Figure 4. Spin-galvanic effect in a Gr-TI heterostructure with the FM easy axis along the y-direction.**
**a,** A colored SEM picture of a representative Gr-TI hybrid device in a Hall bar shape with all ferromagnetic electrodes placed along the y-direction. **b,** A schematic of device 2 and the nonlocal measurement geometry used for probing the SGE (*V*) and reference SV signal (*V*$_2$). **c,** SGE signals ($R = V/I$) measured with Fermi level tuned to the conduction ($V_g$ = 40 V) and valence ($V_g$ = -27 V) bands of proximitized graphene at room temperature. A linear background is subtracted from the data. **d,** Lower panel: The magnitude of SGE signal Δ*R* as a function of gate voltage. The solid line is a guide to the eye. Upper panel: The gate dependence of graphene channel sheet resistance, showing a CNP at $V_g$ = 27 V. **e,** The nonlocal voltage amplitude for the SGE and reference SV signals versus the bias current. Both signals are obtained with the same injector FM contact at *T* = 300 K and $V_g$ = 0 V. Solid lines are guides to the eye. **f,** Normalized amplitude of the SGE signal Δ*R* as a function of the angle *φ* between the magnetic field *B* and the y-axis in the graphene plane. The red line shows the expected trend ~sin(*φ*). The measurements were performed with *T* = 300 K, *I* = -150 μA and $V_g$ = 40 V.

To further investigate the nature of SGE in Gr-TI heterostructures, we also utilized an alternative device design (Fig. 4a), where the FM spin injector contact has the magnetization along the y-direction. This measurement geometry (Fig. 4b) is convenient for SGE measurements with the injected spin polarization aligned persistently in any arbitrary direction. With an application of the external magnetic field along the x-direction ($B_x$), the injected spins experience precession in the y-z plane while the injector FM magnetization **M** gradually rotates in the x-y plane and eventually aligns with the field in the x-direction. Figure 4c shows the measured SGE signal at two gate voltages, where the detected signal $R = V/I$ increases or decreases anti-symmetrically at low fields and saturates as the injector FM rotation completes and spins align in the x-direction at $B_{sat} \approx \pm 0.4$ T. The value of $B_{sat}$ is in good agreement with spin precession experiments (see Supplementary Note 3) and anisotropic magnetoresistance measurements (Supplementary Figure 8).

Similarly to device 1, devices 2 (Fig. 4c) and 3 (Supplementary Figure 9) exhibit a sign change of the SGE signal at the gate voltage $V_0$ that deviates from the average CNP ($V_{CNP}$) in the respective channels. However, while device 1 has similar absolute values of the SGE magnitude below and above $V_0$, in devices 2 and 3 the values on the opposite sides of $V_0$ are rather different. This may happen because the gate voltage, besides tuning the Fermi level in graphene, is predicted to also influence the strength of proximity interaction and the degree of induced Rashba spin splitting in the graphene-based van der Waals heterostructures[30,36]. In addition, the magnitudes of the observed nonlocal SGE signal are significantly larger in devices 2 and 4 (Supplementary Figure 10) compared to devices 1 and 3. These differences between the fabricated devices may arise due to the variability in their interface and relative crystal orientation, which can affect the signal magnitude, SGE efficiency, and position of $V_0$. Nevertheless, the reproducible observation of SGE in both types of devices demonstrates the robustness of the effect and a common spin origin of these spin-charge conversion signals.

Further investigations of the bias dependence of SGE and a reference spin valve signal $V_2$ were performed, as shown in Fig. 4e. Device 2 behaves similarly to device 1, showing an asymmetric voltage trend with a strong correlation between SGE and SV signals that confirms their common origin. Next, the spin-charge conversion process was investigated for spins oriented in different directions by performing the SGE measurements while applying the magnetic field $B_\varphi$ at various angles $\varphi$ in reference to the y-axis in the graphene plane (Supplementary Figure 11). The normalized amplitude of SGE signals $\Delta R$ as a function of angle $\varphi$ follows the expected $\sin(\varphi)$ trend (Fig. 4f), proving the momentum **k** and spin **s** orthogonality of the spin-charge conversion mechanism[32].

Control experiments in the pristine graphene part of the Hall bar structure were performed to rule out stray Hall effect and other spurious charge-based contributions to the observed signal. As our control measurements with the FM injector contact show a null signal in the pristine graphene Hall cross, they prove that the spin-charge conversion signal originates only in the proximitized graphene (Supplementary Figure 12). Further control experiments in separately fabricated Gr-TI Hall bars with all contacts being non-magnetic confirm the absence of SGE-like signals in the devices without a spin-polarized current injection, supporting the spin origin of the observed SGE signal (Supplementary Figure 13). In addition, charge-based contributions are expected to scale linearly with the injection bias current, whereas the measured signal shows a nonlinear behavior typical for spin injection with our tunnel FM electrodes (Fig. 4e, Supplementary Figure 1d).

The efficiency of spin-charge conversion by SGE can be characterized by a unitless parameter $\alpha$, which is estimated from Equation (1):[13]

$$\Delta R_{SGE} = \frac{\alpha P_i R_\square \lambda_s}{W_H} \left( e^{-\frac{L}{\lambda_s}} - e^{-\frac{L+W_H}{\lambda_s}} \right) \qquad (1)$$

, where $\Delta R_{\text{SGE}}$ is the amplitude of the nonlocal SGE signal, $P_{\text{i}}$ is the polarization of the injector contact, $R_\square$ is the graphene sheet resistance, $\lambda_s$ is the spin diffusion length in the Gr-TI hybrid structure, $W_{\text{H}}$ is the width of the Hall bar arms, and $L$ is the distance between the injector and the Hall cross. We obtain the efficiency of 0.17% in device 1, while in devices 2, 3, and 4, it reaches 2.5%, 1%, and 4.8%, respectively, which is comparable to the conversion efficiency seen in Gr-TMD heterostructures[13,14,35]. For comparison with other systems, we can define $\lambda_{\text{IEE}} = \alpha \lambda_s$ yielding an upper bound of $\lambda_{\text{IEE}}$ = 6, 75, 20 and 58 nm in devices 1, 2, 3 and 4 respectively (see Supplementary Note 4). The obtained values are higher than what is seen in heavy metals[37] (0.1-0.4 nm), topological insulators[38] (2.1 nm) and oxide interfaces[39] (6.4 nm), and exceed the $\lambda^*_{\text{ISHE}}$ (<1 nm), a comparative figure of merit for 3D SHE systems[40].

In conclusion, we integrate the 3D topological insulator and 2D graphene in van der Waals heterostructures to demonstrate a gate-tunable spin-galvanic effect, allowing for efficient conversion between spin- and charge-based information at room temperature. Utilizing various device geometries, we observe consistent signals in a spin switch, Hanle spin precession, and magnetization rotation measurement configurations, giving strong evidence of the SGE in the Gr-TI heterostructures. The introduction of graphene mitigates the lack of field effect in strongly doped topological insulators and allows us to probe the energy dependence of their hybrid bands. Importantly, we demonstrate a strong tunability and a sign change of the spin-galvanic signal by the gate electric field, tracing their origin to the spin texture in the Rashba spin-split bands of proximitized graphene. These results open an attractive route for the design of highly tunable spin-orbit phenomena with emerging spin textures based on van der Waals heterostructures, which can become an essential building block in future spintronic circuits and topological quantum technology[41].

Note: After this manuscript was prepared, recent preprints appeared on graphene/TMD heterostructures, including semiconducting $WS_2$ [ref [35]], metallic $TaS_2$ [ref [42]] and 1T'-$MoTe_2$ [ref [43]]. In contrast, our results show a spin-charge conversion effect in a graphene heterostructure with a topologically nontrivial material. We obtain gate-tunable functionality using different types of measurement geometries and spin precession experiments at room temperature, resolving the challenges that were preventing the utilization of TIs in practical spintronic devices.

**Methods**
**Device fabrication**
The van der Waals heterostructures were prepared using CVD Gr (from Grolltex Inc) on highly doped Si (with a thermally grown 285-nm-thick $SiO_2$ layer). The BST flakes (single crystals grown from a melt using a high vertical Bridgeman method, acquired from Miracrys) were exfoliated by conventional scotch tape technique and dry-transferred on top of Gr inside a glovebox. The exfoliation and heterostructure preparation in a controlled $N_2$ environment is expected to provide higher quality Gr-TI interfaces. Next, appropriate 90-140 nm-thick TI flakes were identified by optical microscopy for device fabrication. The Gr and the Gr-TI heterostructure channels were patterned in Hall bar shapes by electron beam lithography (EBL), followed by oxygen plasma etching. The non-magnetic and magnetic contacts were patterned on Gr in two subsequent EBL steps. The Ti/Au contacts were employed as non-magnetic electrodes on the Gr.

For the preparation of ferromagnetic contacts, we used electron beam evaporation to deposit 0.6 nm of Ti, followed by *in situ* oxidation in a pure oxygen atmosphere for 10 minutes to form a $TiO_2$ tunneling barrier layer. Without exposing the device to the ambient atmosphere, in the same chamber, we deposited 90 nm of Co, after which the devices were finalized by liftoff in warm acetone at 65°C. In the final devices, the Co/$TiO_2$ contacts on Gr act as the source for spin-polarized electrons, the Gr-TI heterostructure region serves as a channel, and the $n^{++}$ Si/$SiO_2$ is used as a back gate. The FM tunnel contact resistances, measured in a three-terminal configuration, were around 1-3 kΩ. The field-effect mobility of the Gr channel in proximity with TIs is ~1600 $cm^2V^{-1}s^{-1}$. To estimate the interface resistance of Gr-BST in the heterostructure, the graphene stripes were first patterned, followed by exfoliation of BST inside the glovebox and evaporation of contacts on both the Gr and BST. The Gr-BST interface resistance is ~ 3 kΩ at zero-bias conditions (see Supplementary Figure 4).

**Electrical measurements**

The spin-galvanic effect measurements were performed in a cryostat with the sample rotation stage and variable magnetic field. The sample was kept in vacuum, and the measurements were performed at room temperature and at $T$ = 40 K. In the experiments, a spin injection bias current was applied using a Keithley 6221 current source, and the nonlocal voltage was detected across the Hall bar structure of Gr-BST using a Keithley 2182A nanovoltmeter; the gate voltage was applied with the $SiO_2$/Si back gate using a Keithley 2400 multimeter.

**Data availability**

The data that support the findings of this study are available from the corresponding authors on a reasonable request.

**Acknowledgments**

The authors acknowledge financial support from the European Union's Horizon 2020 research and innovation program under grant agreements no. 696656, no. 785219, and no. 881603 (Graphene Flagship Core 1, Core 2, and Core 3), Swedish Research Council VR project grants (No. 2016-03658), EU FlagEra project (funded by Swedish Research Council VR No. 2015-06813), VINNOVA 2D Tech competence center, Graphene center, and the EI Nano program at Chalmers University of Technology. We acknowledge useful discussions with Bing Zhao in our group. We also acknowledge the support from staff at Quantum Device Physics Laboratory and Nanofabrication laboratory at Chalmers University of Technology.

**Author Contributions**
SPD and DK conceived the idea and designed the experiments. DK fabricated and characterized the devices. BK and AMH participated in device fabrication and discussions. DK and SPD analyzed and interpreted the experimental data, compiled the figures, and wrote the manuscript. SPD supervised the research project.

**Competing Interests**
The authors declare no competing interests.

**Corresponding authors**:
Correspondence and requests for materials should be addressed to Saroj P. Dash, Email: saroj.dash@chalmers.se

# Supplementary Information

# Gate-tunable Spin-Galvanic Effect in Graphene-Topological Insulator van der Waals Heterostructures at Room Temperature

Dmitrii Khokhriakov[1], Anamul Md. Hoque[1], Bogdan Karpiak[1], Saroj P. Dash[1*]

[1]Department of Microtechnology and Nanoscience, Chalmers University of Technology, SE-41296, Göteborg, Sweden

## Supplementary Note 1

**Bias dependence of spin-galvanic and reference signals in Gr-TI heterostructures**

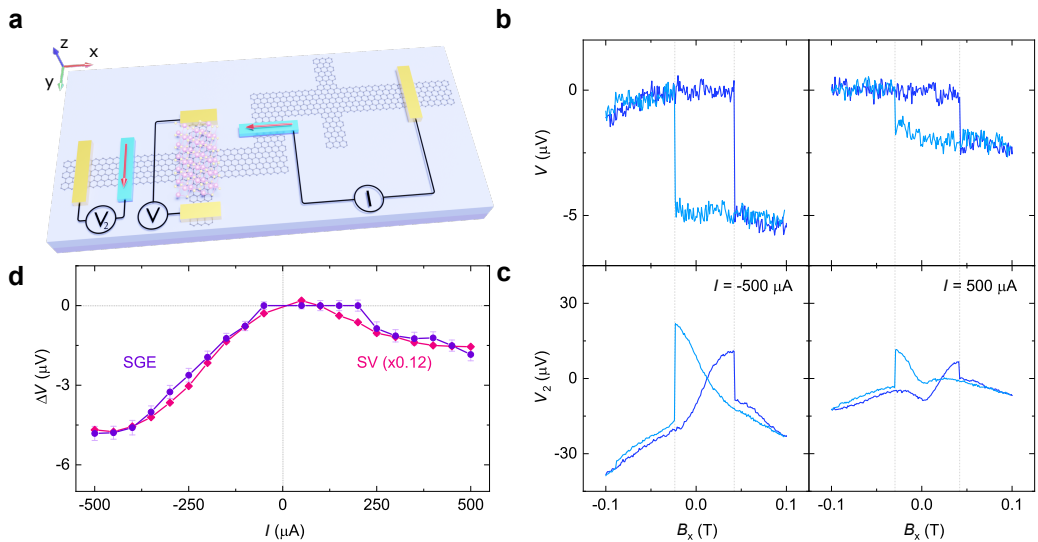

**Supplementary Figure 1. Bias dependence of the spin-galvanic signal measured via spin switch in device 1. a,** A schematic of the device and the measurement geometry. **b,** SGE signals obtained with the bias current $I$ = ±500 µA. **c,** Reference "spin valve" signals measured simultaneously to the data shown in (b). **d,** SGE signal amplitude as a function of the bias current $I$ plotted together with the amplitude of "spin valve" switches. The amplitude of the reference signal is scaled by a factor of 0.12, emphasizing the strong correlation in the trends of both signals. The measurements were performed at $T$ = 40 K and $V_g$ = -80 V.

To confirm the relation between the spin-galvanic effect (SGE) and spin transport in the device, we performed measurements at various values of the applied bias current $I$ to the injector ferromagnet (Supplementary Figure 1a). First, the SGE voltage $V$ and the reference spin voltage $V_2$ were recorded while sweeping the magnetic field aligned in the x-direction (Supplementary Figures 1b,c). The signal in the reference geometry shows a combined behavior of a spin valve-like switching, originating from the injector ferromagnet that is collinear to the $B_x$, and an xHanle-like continuous rotation of the detector that is orthogonal to the $B_x$. Therefore, the jump amplitude in $V_2$ shows not the full spin signal magnitude but only its projection onto the x-component of the detector magnetization, which is small due to the relatively low field at which the switching occurs. Nevertheless, the magnitude of these jumps fairly represents the trend of the spin signal with bias current, while their positions correlate well with those observed in SGE, demonstrating their common origin. The complete trends of the SGE spin switch magnitude and (scaled) $V_2$ jump amplitude with the bias current are shown in Supplementary Figure 1d. Both signals have an asymmetric dependence on the bias $I$, where switching occurs in the same direction for both +$I$ and -$I$. Such behavior is commonly observed for spin transport in graphene with tunneling contacts and is generally assigned to magnetic proximity effects and energy-dependent spin-resolved density of states at the injector FM/Gr interface[1]. A strong correlation between the bias trends of the SGE and the reference spin signal establishes their common spin-based origin.

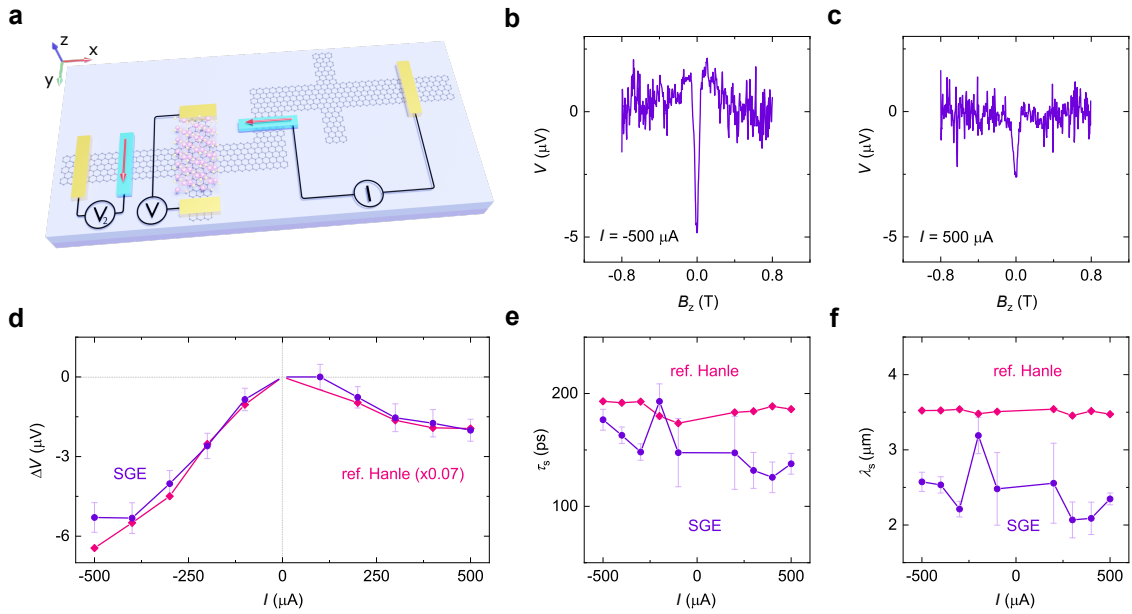

**Supplementary Figure 2. Bias dependence of the spin-galvanic signal measured via spin precession in device 1. a,** A schematic of the device and the measurement geometry. **b,c,** SGE signals obtained with the bias current $I$ = ±500 µA. **d,** SGE signal amplitude as a function of the bias current $I$ plotted together with the amplitude of reference Hanle curves measured by $V_2$. The amplitude of the reference signal is scaled by a factor of 0.07 to emphasize the good correlation in the trends of both signals. **e,f,** Spin lifetime $\tau_s$ and spin diffusion length $\lambda_s$ extracted by fitting the Hanle equation to SGE and also to reference Hanle signals. The measurements were performed at $T$ = 40 K and $V_g$ = -80 V.

Similar results were also obtained when measuring the SGE via spin precession in $B_z$, as shown in Supplementary Figure 2. However, unlike the spin switch, which only allows us to assess the magnitude of the spin signal, these measurements allow us to extract spin transport parameters of the system by fitting the curves to the classic Hanle equation (Supplementary Equation 1),

$$V(B) = Re\left\{\frac{P_i P_d I R_\square \widetilde{\lambda}_s}{W_{gr}} e^{-\frac{L}{\widetilde{\lambda}_s}}\right\} \quad (1)$$

with

$$\widetilde{\lambda}_s = \frac{\lambda_s}{\sqrt{1+i\omega_L \tau_s}} \quad (2)$$

where $P_i$ and $P_d$ are the spin polarization values for the injector and detector contacts, $R_\square$ is the sheet resistivity of graphene, $W_{gr}$ is the graphene channel width, $\omega_L = \frac{g\mu_B}{\hbar} B$ is the Larmor spin precession frequency, $\mu_B$ is Bohr magneton, $L$ is the channel length, $\lambda_s = \sqrt{D_s \tau_s}$ is the spin diffusion length with $D_s$ and $\tau_s$ being the spin diffusion coefficient and spin lifetime, respectively. If the injector and detector contacts are placed perpendicularly, as is the case for the reference Hanle measurements, an imaginary part is used in Supplementary Equation 1. From a single measurement, $P_i$ and $P_d$ cannot be extracted independently, and the average spin polarization of the ferromagnetic contacts is calculated $P = \sqrt{P_i P_d}$. Supplementary Figures 2e,f show the values of $\tau_s$ and $\lambda_s$ extracted from both the symmetric SGE Hanle signal and antisymmetric reference Hanle signal. While the uncertainty in the extracted values is higher for the SGE due to its smaller signal-to-noise ratio, a good correspondence between the parameters demonstrates the validity of the SGE in Gr-TI heterostructures. It allows for their applications in all-electrical spintronic devices, where spin polarization can be created and detected without the use of ferromagnets.

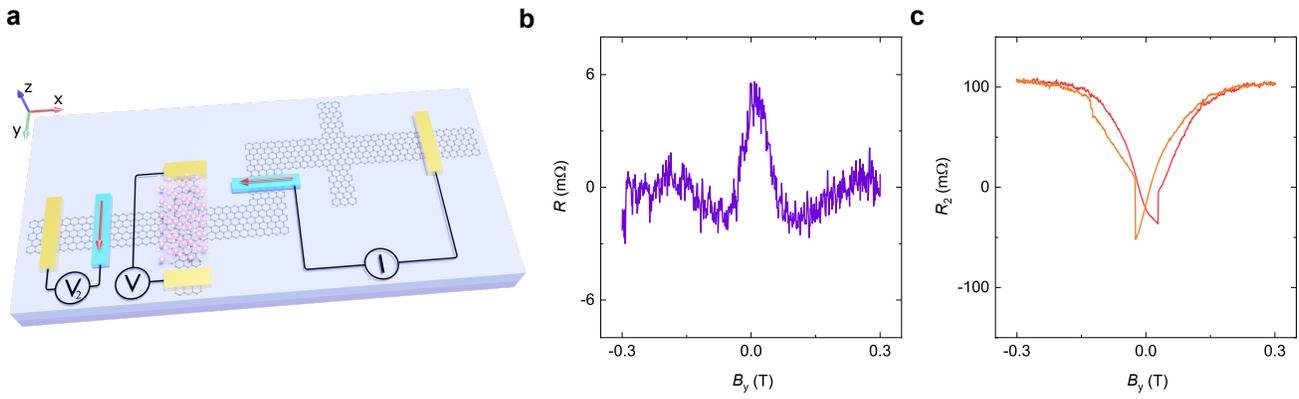

**Supplementary Figure 3. SGE spin precession measurement obtained with B$_y$ sweep. a,** A schematic of the device and the measurement geometry. **b**, SGE measurement $R = V/I$ obtained by sweeping the magnetic field along the y-direction, B$_y$. A symmetric SGE signal is observed, whereas no antisymmetric ISHE contribution from $s_z$ spins is visible. **c,** Reference spin transport signal $R_2 = V_2/I$ measured simultaneously with the data shown in (b). The signal shows a mixture of the sharp switching of the detector electrode magnetization and a gradual rotation of the injector by the magnetic field. All the measurements were performed at $T$ = 300 K with $I$ = -500 µA and $V_g$ = -80 V.

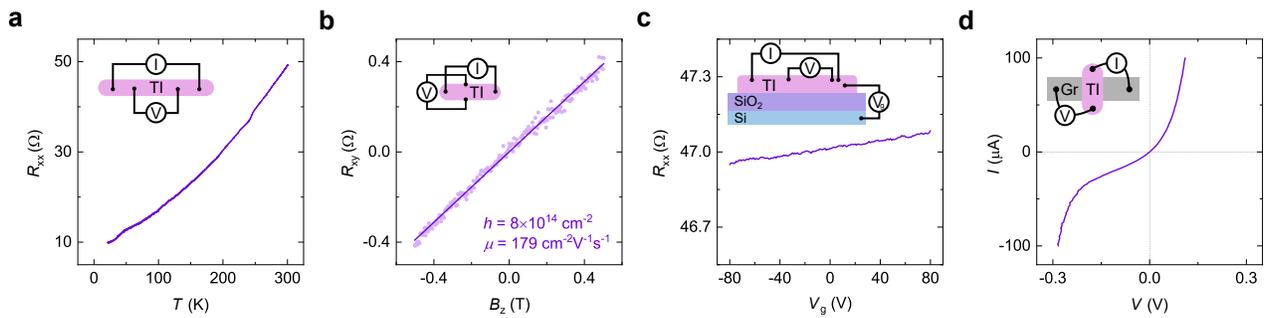

**Supplementary Figure 4. Characterization of the topological insulator BST. a,** Four-probe resistance of a BST flake as a function of temperature, showing a metallic behavior of the sample. **b,** Room temperature Hall effect measurements on a BST flake show hole doping (p-type) of the TI. **c,** Resistance of a BST flake as a function of the gate voltage. Only a marginal change in resistance is observed, illustrating the suppressed field effect in the TI due to its high doping. **d,** A current-voltage (IV) characteristic of the Gr-TI interface in a four-terminal measurement geometry shows a non-linear behavior with zero-bias interface resistance of around 3.2 kΩ at room temperature. Measurements were carried out in a device where the TI and graphene are contacted by Ti/Au electrodes.

## Supplementary Note 2

### Gate tunability of the Gr and TI spin textures

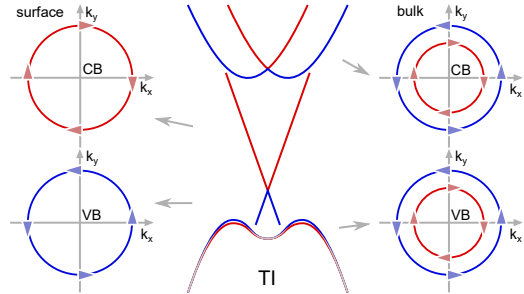

**Supplementary Figure 5. Band structure of a TI.** A schematic of the simplified band structure of the TI surface and Rashba-split bulk states with corresponding Fermi contours indicating their spin textures. The colors represent the helicity of the bands.

The band structure of the TI (Supplementary Figure 5) is relatively complex, with opposite signs of the surface and bulk spin textures in the conduction band and an asymmetric convoluted dispersion near the valence band edge[2–6]. Therefore, we evaluate the possibility of additional contributions to the observed signal from spin-charge conversion processes in the TI. Specifically, we consider the SGE in the TI surface states, as well as SGE and an inverse spin hall effect (ISHE) in the TI bulk bands. As the TI surface spin texture is opposite for holes and electrons, SGE in these states it is not expected to produce a sign change in the measurement[4,5]. On the other hand, the spin-charge conversion processes in the TI bulk bands could produce a sign change of the signal, if the Fermi level ($E_f$) in the TI could be tuned across its bandgap. However, since the TI is highly doped, its Fermi level is unlikely to experience such large tunability, which is confirmed by a rather small change in the TI resistance as a function of the gate voltage (see Supplementary Figure 4c). Thus, we consider the $E_f$ in TI to be fixed, and therefore none of its contributions should produce a sign change in the measurement. However, if several contributions are present, of which some change sign at the CNP (e.g. SGE in graphene) and some do not (all in the TI), the sign change point ($V_0$) in the resulting signal and its relative position to the graphene CNP ($V_{CNP}$) can vary depending on the relative sign and efficiency of each contributing process.

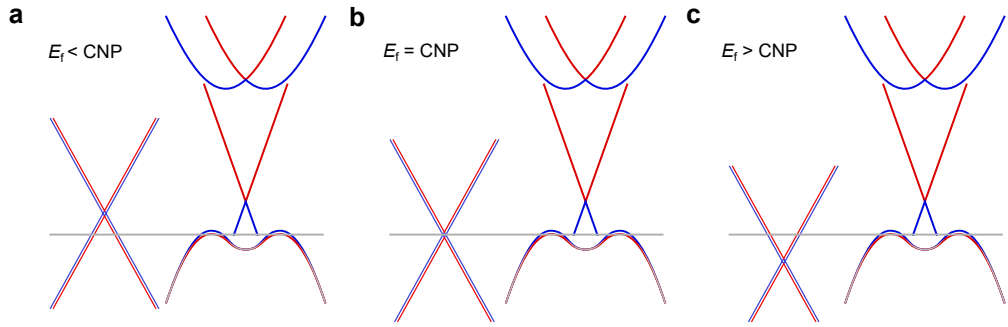

**Supplementary Figure 6. Gate tunability of the bands in a Gr-TI heterostructure.** The band alignment diagram in a Gr-TI heterostructure, where the gate voltage shifts the Fermi level in graphene but not in the TI. The panels correspond to (**a**) $E_f <$ CNP, (**b**) $E_f =$ CNP, (**c**) $E_f >$ CNP.

Supplementary Figure 6 shows three cases of the band alignment in a Gr-TI heterostructure, with the graphene Fermi level being tuned below, at or above its CNP. Taking into account the doping in each band and its helicity, we summarize the expected contribution from each described process. For simplicity, we assume the injected spins to be polarized in the positive y-direction, and record in Supplementary Table 1 the expected carrier type (h for holes and e for electrons) and the sign of their acquired momentum along the x-direction.

Supplementary Table 1. **Expected spin-charge conversion signs in the Gr and TI bands.**

|  | SGE in Gr | SGE in TI surface | SGE in TI bulk | ISHE in TI bulk |
|---|---|---|---|---|
| $E_f <$ CNP$_{Gr}$ | h+ | h+ | h+ | h+ |
| $E_f =$ CNP$_{Gr}$ | 0 | h+ | h+ | h+ |
| $E_f >$ CNP$_{Gr}$ | e+ | h+ | h+ | h+ |

From Supplementary Table 1 one can see that, when graphene is tuned to the valence band ($V_g < V_{CNP}$ in our experiment), all contributions have the same sign, whereas the proximitized graphene gives an opposite sign to the TI when it is tuned into the conduction band. Thus, a shift of the sign change point from the CNP to the higher energies (higher $V_g$) can be expected, whereas, experimentally, $V_0$ is below $V_{CNP}$. Therefore, the observed shift is likely to have a different origin, possibly a local CNP in the Gr-TI heterostructure region.

# Supplementary Note 3

## Spin transport in graphene and Gr-TI heterostructures

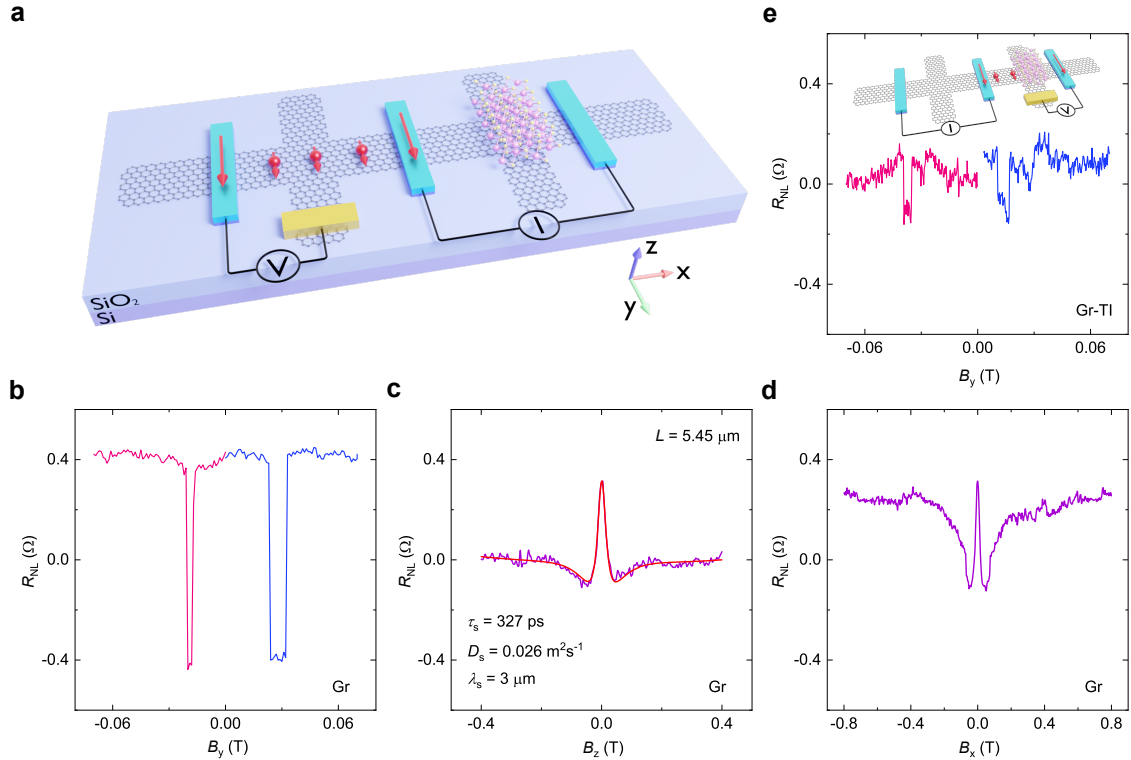

**Supplementary Figure 7. Spin transport in the pristine graphene and Gr-TI channels of device 2. a**, A schematic of the device and measurement geometry used to characterize the spin transport in graphene. **b**, A spin valve signal observed in the pristine graphene region with the measurement geometry shown in (a). **c,d,** Corresponding Hanle spin precession signals with magnetic fields applied in z and x directions, respectively. **e**, A spin valve signal measured in the Gr-TI heterostructure region. The inset shows the utilized measurement geometry.

Spin transport in graphene was characterized using a conventional nonlocal spin valve and Hanle spin precession experiments. The spin accumulation in graphene is created by passing a current $I$ between the injector ferromagnetic contact and graphene, while the nonlocal (NL) spin-dependent voltage $V_{\text{NL}}$ is detected by another ferromagnetic contact placed at a length $L$ away from the injector with reference to a spin-insensitive electrode, as indicated in Supplementary Figure 7a. To perform the spin-valve measurement, we sweep the in-plane magnetic field $B_y$ along the easy axis of the ferromagnetic contacts while recording the nonlocal resistance $R_{\text{NL}} = V_{\text{NL}}/I$. Sharp changes in $R_{\text{NL}}$ are measured when the magnetization of the injector or detector switches giving either parallel or antiparallel configuration, as shown in Supplementary Figure 7b.

Hanle spin precession measurements were performed by measuring the nonlocal resistance $R_{NL}$ while sweeping an out-of-plane magnetic field $B_z$, which induces spin precession and dephasing. From fitting the data (Supplementary Figure 7c) with the Supplementary Equation 1, we extract $\Delta R_{NL} \sim 0.33$ Ω, $\tau_s$ = 327 ps, $D_s$ = 0.026 m²s⁻¹ and $\lambda_s = \sqrt{D_s \tau_s}$ = 3 µm, with the channel length of $L$ = 5.45 µm.

An application of the magnetic field along the x-axis induces spin precession in the graphene channel, while further leading to the rotation of the ferromagnetic contact magnetization. The data measured in this geometry are shown in Supplementary Figure 7d. The saturation of the observed signal occurs at fields $|B| > B_{sat} \approx 0.4$ T, which indicates the fully rotated contact magnetization along the x-axis.

After the presence of spin transport in the pristine graphene was established, we performed similar experiments in the channel having the Gr-TI heterostructure. Supplementary Figure 7e shows the obtained spin-valve data with the measurement geometry depicted in the inset. Compared to the pristine Gr channel, the spin signal in the heterostructure is significantly reduced. This behavior can be due to the modified graphene properties caused by the proximity-induced SOC[7].

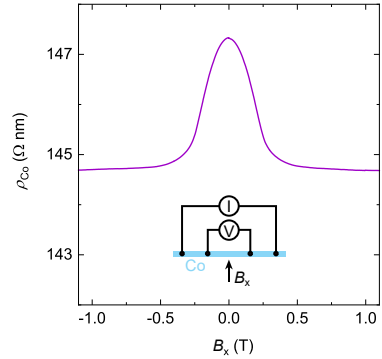

**Supplementary Figure 8. Anisotropic magnetoresistance of the ferromagnetic Co electrode.** The resistivity of a Co contact as a function of the external magnetic field applied along the x-axis indicates a gradual rotation of magnetization and its alignment with the external magnetic field beyond the saturation field value $B_{sat} \approx \pm 0.4$ T. The inset shows the utilized four-probe measurement geometry.

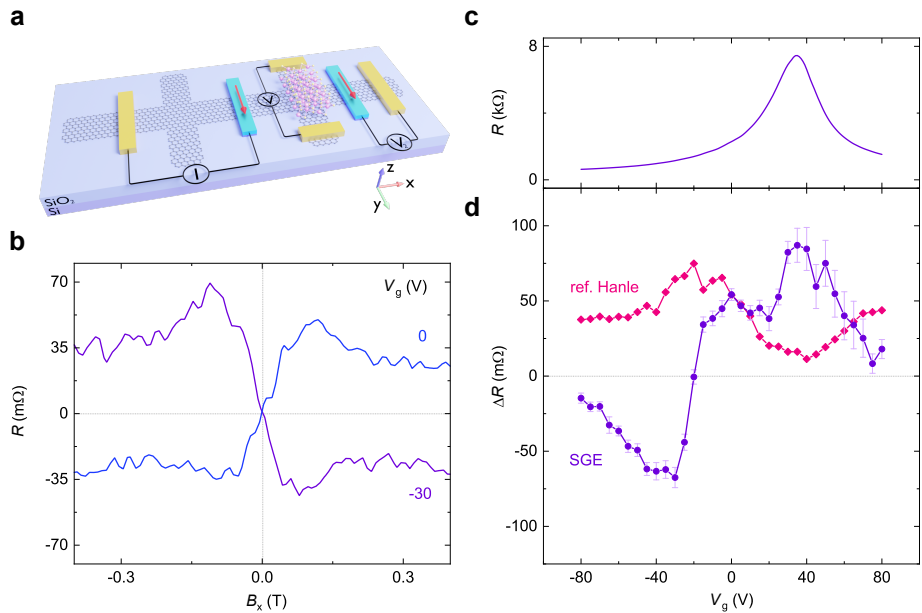

**Supplementary Figure 9. Spin-galvanic effect in the Gr-TI heterostructure device 3 with the FM easy axis along the y-direction. a,** A schematic of the device and the nonlocal measurement geometry used for probing the SGE ($R = V/I$) and reference Hanle ($R_2 = V_2/I$). **b,** SGE signals measured at different gate voltages across the sign change point $V_0 = -20$ V. A linear background is subtracted from the data. The saturation field is smaller compared to device 2 because of different thickness of the FM contacts. **c,** The gate dependence of graphene channel resistance, showing a CNP at $V_g = 34$ V. **d,** The magnitude of SGE and reference Hanle signals as a function of the gate voltage. The measurements were performed with $I = -500$ µA at $T = 40$ K.

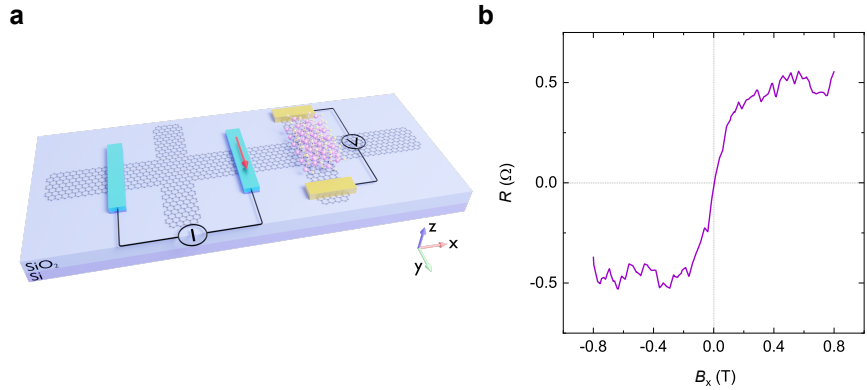

**Supplementary Figure 10. Spin-galvanic effect in the Gr-TI heterostructure device 4 at room temperature. a,** A device schematic with the SGE measurement geometry. **b,** A spin-galvanic signal $R = V/I$ measured in the Gr-TI heterostructure of the device 4 at $I$ = -200 µA and gate voltage $V_g$ = -10 V at room temperature.

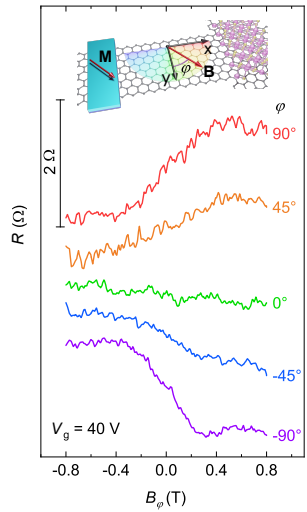

**Supplementary Figure 11. Angle dependence of the SGE signals in device 2.** SGE signals measured at various angles $\varphi$ with $I$ = -150 µA and $V_g$ = 40 V. The inset shows the utilized coordinate axes with the angle $\varphi$ between the magnetic field $B$ and the y-axis in the graphene plane. The angle $\varphi$ = +/-90° corresponds to the positive $B$ field applied in +/-x directions. A linear background was subtracted from the data. As the magnetization direction of the FM injector is controlled by the magnetic field, these measurements allow to study the spin-charge conversion for spin polarization oriented persistently in any in-plane direction. The obtained signal shows a strong dependence on the orientation of the magnetic field conforming to the expected sin($\varphi$) trend (see Fig. 4f in the main text), consistent with the orthogonality between the momentum **k** and spin **s** required for spin-charge conversion via SGE.

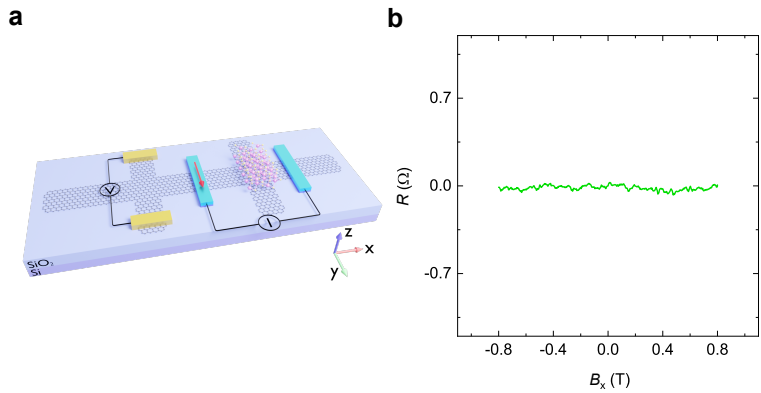

**Supplementary Figure 12. Control experiments in the pristine Gr Hall cross of the device 2. a,** A schematic showing the SGE measurement geometry used to obtain the reference data in the pristine graphene Hall cross. **b,** A signal measured in the SGE geometry in a reference graphene Hall cross of the device 2 with $I$ = -40 µA, $T$ = 300 K and $V_g$ = 0 V. The null signal in the pristine graphene confirms that an increased SOI is required to observe spin-charge conversion by SGE.

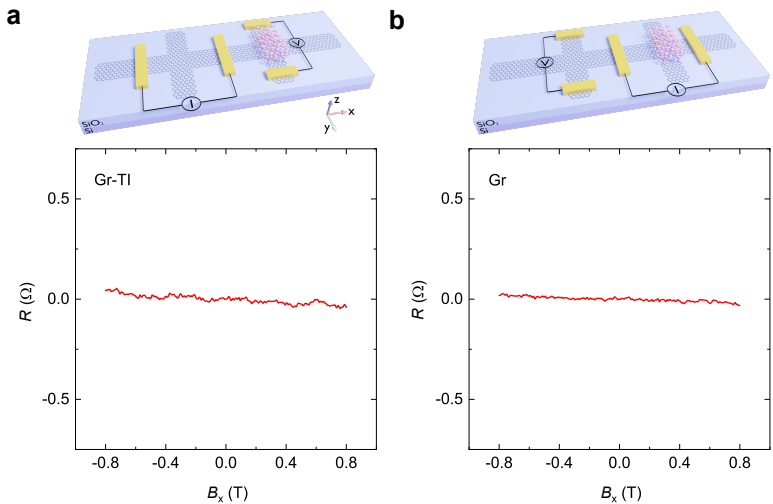

**Supplementary Figure 13. Control experiments in Gr-TI heterostructure and pristine Gr areas of a Hall bar with non-magnetic Ti/Au contacts. a,** A measurement schematic, and the obtained data with the Ti/Au injector and detector contacts in the SGE measurement geometry within the Gr-TI heterostructure Hall cross. **b,** Similar measurements in the pristine graphene Hall cross with $I$ = -200 µA, $T$ = 300 K and $V_g$ = 0 V. The absence of an SGE-like signal here proves that no spurious charge-based effects contribute to the measurements. The observed linear background may be caused by a normal Hall effect due to the presence of an unintentional out-of-plane magnetic field component, which can appear due to a small deviation of chip alignment because of the limited accuracy of a sample rotator.

# Supplementary Note 4

## Quantification of the Rashba-Edelstein effect in graphene

We follow the approach described in ref. 8 to quantify the efficiency of spin-charge conversion via spin-galvanic effect (same as proximity-induced inverse Rashba-Edelstein effect) in graphene. Assuming the spin direction to be orthogonal to the electron momentum, the SGE signal magnitude can be written as

$$\Delta R_{\text{SGE}} = \frac{eP_i\gamma R_\square^2\lambda_s^2}{W_H}\left(e^{-\left(\frac{L}{\lambda_s}\right)} - e^{-\left(\frac{L+W_H}{\lambda_s}\right)}\right) \tag{3}$$

, where $\gamma$ is the conversion efficiency between the spin accumulation and charge current density, $W_H$ is the width of Hall bar arms, and $P_i$ is the polarization of the ferromagnetic spin injector contact. For a direct comparison with the performance of FM contacts, one can rewrite Supplementary Equation 3 introducing a unitless parameter $\alpha = e\gamma R_\square \lambda_s$ that describes the efficiency of spin-charge conversion by SGE:

$$\Delta R_{\text{SGE}} = \frac{\alpha P_i R_\square \lambda_s}{W_H}\left(e^{-\left(\frac{L}{\lambda_s}\right)} - e^{-\left(\frac{L+W_H}{\lambda_s}\right)}\right) \tag{4}$$

This equation is analogous to Supplementary Equation 1 with $\alpha$ acting as a detector spin polarization $P_d$, allowing for a direct comparison between these different detection mechanisms. Supplementary Table 2 summarizes the parameters of spin transport and spin-charge conversion in our devices.

Supplementary Table 2. **The extracted spin transport and SGE parameters.**

|  | $P$ (%) | $R_\square$ (Ω) | $\lambda_s$ (μm) | $W_H$ (μm) | $W_{\text{gr}}$ (μm) | $L$ (μm) | $\Delta R_{\text{SGE}}$ (Ω) | $\alpha$ (%) | $\gamma$ (AJ$^{-1}$m$^{-1}$) | $\lambda_{\text{IEE}}$ (nm) |
|---|---|---|---|---|---|---|---|---|---|---|
| Device 1 | 6.7 | 446 | 3.5 | 1.3 | 2.6 | 4.1 | 0.013 | 0.17 | 6.7x10$^{18}$ | 6 |
| Device 2 | 7.4 | 595 | 3 | 2.1 | 2.8 | 1.9 | 0.43 | 2.5 | 8.8x10$^{19}$ | 75 |
| Device 3 | 9 | 453 | 2 | 1.8 | 4 | 3 | 0.062 | 1 | 7.2x10$^{19}$ | 20 |
| Device 4 | 7.4 | 887 | 1.21 | 0.75 | 1.85 | 1.1 | 0.94 | 4.8 | 2.8x10$^{20}$ | 58 |

We would like to note that the inverse Edelstein effect is often characterized by the IEE length $\lambda_{\text{IEE}} = I_c^{2D}/I_s^{3D}$ that describes the conversion efficiency from 3D spin current to 2D charge current and therefore has a unit of length[9]. Since in our measurement geometry the spin current injection and charge current detection both happen in 2D graphene, these currents have the same dimensionality and the corresponding efficiency merit $\alpha$ is unitless. For the sake of comparison, we can define $\lambda_{\text{IEE}} = \alpha \lambda_s$ yielding $\lambda_{\text{IEE}}$ = 6, 75, 20 and 58 nm in devices 1, 2, 3 and 4 respectively. The obtained values are higher than what is seen in heavy metals[10] (0.1-0.4 nm), topological insulators[11] (2.1 nm) and oxide interfaces[12] (6.4 nm), and exceed the $\lambda_{\text{ISHE}}^*$ (<1nm), a comparative figure of merit for 3D SHE systems[13]. However, the obtained $\lambda_{\text{IEE}}$ can only be used for qualitative comparison, whereas $\alpha$ is the proper merit of spin-charge conversion efficiency by SGE in a 2D system such as graphene in proximity to a TI.

**Supplementary References**